\documentclass[reprint, pra, a4paper, aps, 10pt, superscriptaddress, amsmath, amssymb, twocolumn, showkeys, floatfix]{revtex4-2}
\usepackage{hyperref}
\usepackage{graphicx}
\usepackage{array}
\usepackage{multirow}
\usepackage{makecell}
\usepackage[table]{xcolor}
\usepackage{hhline}
\usepackage{booktabs}
\usepackage{tabularx}
\usepackage{bm}
\usepackage{braket}

\begin{document}

\title{Experimental Demonstration of Free-Space Unidimensional Continuous-Variable Quantum Key Distribution Under High Detector Noise}

\author{Rachita Nandan} \email{rachita@prl.res.in}
\affiliation{Quantum Science and Technology Laboratory, Physical Research Laboratory, Ahmedabad, India 380009.}
\affiliation{Indian Institute of Technology, Gandhinagar, India 382355.}

\author{Jayanth Ramakrishnan}
\affiliation{Quantum Science and Technology Laboratory, Physical Research Laboratory, Ahmedabad, India 380009.}

\author{Shashi Prabhakar}
\affiliation{Quantum Science and Technology Laboratory, Physical Research Laboratory, Ahmedabad, India 380009.}

\author{R. P. Singh}
\affiliation{Quantum Science and Technology Laboratory, Physical Research Laboratory, Ahmedabad, India 380009.}

\date{\today}

\begin{abstract}
Continuous-variable quantum key distribution (CV-QKD), which uses quadratures of the electromagnetic field, enables practical quantum communication using standard telecommunication technologies. Unidimensional CV-QKD (UD-CVQKD) simplifies the implementation by restricting modulation to a single quadrature. In this work, we experimentally demonstrate a free-space Gaussian-modulated UD-CVQKD system operating under a high detector electronic-noise regime ($1.4$ shot-noise units). The system employs polarized coherent states with signal and local oscillator co-propagating in the same spatial mode in orthogonal polarizations, ensuring stable interference. System security is analyzed under both untrusted (UTD) and trusted (TD) detector noise models. While no positive secret key rate is obtained under the UTD model, the TD model enables secure key generation over a finite range of modulation variances, highlighting the critical role of detector trust in high-noise conditions. A maximum secret key rate of $270$ kbps is achieved at an optimal modulation variance of $11.57$. Furthermore, secure operation requires high-transmittance (low-loss) channels under such noise conditions. This study demonstrates the practical feasibility of free-space UD-CVQKD in realistic high electronic-noise detection constraints and highlights detector electronic noise as a key limiting factor in practical systems.   

\end{abstract}


\maketitle

\section{Introduction} \label{sec:introduction}

Quantum key distribution (QKD) enables two distant parties to share secret cryptographic keys with information-theoretic security guaranteed by the principles of quantum mechanics \cite{bennett2014quantum, scarani2009security, pirandola2020advances}. Practical QKD systems have been realized over both fiber and free-space links, with long-distance implementations largely demonstrated using discrete-variable QKD (DV-QKD) \cite{ribordy2000long, liao2017satellite}, where information is encoded in discrete quantum states. Unlike DV-QKD, continuous-variable QKD (CV-QKD) encodes information in the continuous quadratures of the electromagnetic field and is measured using homodyne or heterodyne detection \cite{grosshans2002continuous, ralph1999continuous}. Owing to its compatibility with standard telecommunication components \cite{weedbrook2012gaussian, zhang2024continuous, pirandola2021composable}, high detection efficiencies \cite{zhang20181, huang2013300}, and potential for high secret key rates over metropolitan distances \cite{wang2018high, schrenk2017high}, CV-QKD is considered a promising approach for practical quantum communication. 

The Gaussian-modulated coherent-state (GMCS) CV-QKD protocols, notably the GG02 protocol, employ symmetric modulation of both amplitude and phase quadratures \cite{grosshans2003quantum, lodewyck2007quantum, qi2007experimental} and have well-established security proofs \cite{garcia2006unconditional, leverrier2010finite, leverrier2015composable}. To reduce implementation complexity, asymmetric modulation schemes have been proposed, where information is encoded in a single quadrature, referred to as unidimensional CV-QKD (UD-CVQKD) \cite{usenko2015unidimensional}. This approach simplifies both state preparation and detection, offering a simplified and resource-efficient implementation compared to the standard GG02 protocol. Despite the asymmetric modulation, security can be ensured by appropriately constraining the parameters of the unmodulated quadrature, with both finite-size and composable security established \cite{wang2017finite, liao2018composable}. UD-CVQKD protocols have also been experimentally demonstrated in both fiber and free-space channels \cite{wang2017experimental, shen2019free}. Although it may exhibit reduced performance under high channel loss compared to GG02, it remains comparable in low-loss channels \cite{shen2019free}. This reflects a trade-off between implementation simplicity and achievable performance. 

In practical CV-QKD systems, homodyne (or heterodyne) detection requires high interferometric visibility and a stable phase reference between the signal and the local oscillator (LO), which is challenging to maintain in phase-sensitive free-space channels. Polarization-based encoding \cite{lorenz2004continuous, lorenz2006witnessing} provides an effective alternative, as free-space channels largely preserve polarization, eliminating the need for active phase locking and thus simplifying the receiver design. Moreover, co-propagation of the signal and LO in the same spatial mode ensures that both experience similar atmospheric disturbances, leading to partial auto-compensation of phase fluctuations \cite{elser2009feasibility, heim2010atmospheric, heim2014atmospheric} and improved interferometric stability. Leveraging these advantages, in this work we experimentally realize a free-space UD-CVQKD system using coherent polarization states, where the signal and LO co-propagate in a single spatial mode with orthogonal polarizations.      

Detection noise plays a critical role in practical CV-QKD implementations and must be properly modeled \cite{pan2025detector}. The treatment of detector noise as trusted or untrusted directly influences security analysis. While untrusted models provide conservative bounds at the cost of reduced performance, trusted models improve key rates by distinguishing intrinsic detector noise from potential adversaries, requiring detector characterization by the legitimate parties. In practice, commercially available detectors exhibit intrinsic imperfections, including finite detection efficiency and significant electronic noise, which can degrade system performance and even compromise security, particularly when treated as untrusted. A previous study has explored classical optical preamplification techniques \cite{fossier2009improvement} to mitigate the limitations of finite detection efficiency. In this work, we investigate the performance and security of a free-space GMCS UD-CVQKD system operating under realistic high detector electronic-noise conditions, corresponding to a low shot-noise to electronic-noise clearance of approximately 2.4 dB. We present an experimental demonstration supported by theoretical security analysis and numerical simulations, and systematically evaluate system performance under both trusted and untrusted detector noise models. We further analyze the dependence of key security parameters on Alice’s modulation variance and compare experimental results with numerical predictions. Our results provide a practical assessment of UD-CVQKD in high detector electronic-noise regimes, identifying the operational limits for secure key generation and highlighting the requirement of high-transmission channels under such conditions.

This article is organized as follows. Section~\ref{sec:theory} presents the theoretical background relevant to this work. Section~\ref{sec:simulation} provides the numerical security analysis and performance simulations of the protocol. The experimental implementation of the UD-CVQKD system is detailed in Section~\ref{sec:experimentalsetup}. The corresponding results and their analysis are discussed in Section~\ref{sec:rnd}. Finally, Section~\ref{sec:conclusion} summarizes the findings and concludes the study.

\section{Theoretical Framework} \label{sec:theory}

This section presents the system model and theoretical framework of the UD-CVQKD protocol. Polarization-based encoding using coherent states is described in \ref{subsec:encoding}. The protocol execution is described in \ref{subsec:protocol}, followed by the formulation of the considered detector noise models (trusted and untrusted) in \ref{subsec:noisemodels}. 

\subsection{Polarization-Based Encoding}\label{subsec:encoding}

Information is encoded in polarized coherent states \cite{lorenz2004continuous}, which are conveniently described using the quantum Stokes operators \cite{korolkova2002polarization}
\begin{align}
    \hat{S}_0 &= \hat{a}_H^\dagger \hat{a}_H + \hat{a}_V^\dagger \hat{a}_V, \quad
    \hat{S}_1 = \hat{a}_H^\dagger \hat{a}_H - \hat{a}_V^\dagger \hat{a}_V, \nonumber \\
    \hat{S}_2 &= \hat{a}_H^\dagger \hat{a}_V + \hat{a}_V^\dagger \hat{a}_H, \quad
    \hat{S}_3 = i\left(\hat{a}_V^\dagger \hat{a}_H - \hat{a}_H^\dagger \hat{a}_V\right),
\end{align}
where $H$ and $V$ denote the horizontal and vertical polarization modes, and $\hat{a}^\dagger$ and $\hat{a}$ are the corresponding creation and annihilation operators. The Stokes operators satisfy the commutation relations $[\hat{S}_j, \hat{S}_k] = 2i \,\epsilon_{jkl} \hat{S}_l$, with $j,k,l \in \{1,2,3\}$, and therefore obey uncertainty relations analogous to those of the quadrature operators $\hat{X}$ and $\hat{P}$, implying that two Stokes operators cannot be simultaneously measured with arbitrary precision when the third has a nonzero expectation value. The corresponding uncertainty relation is given by 
\begin{equation}
    \mathrm{Var}(\hat{S}_2)\,\mathrm{Var}(\hat{S}_3) \geq \left| \langle \hat{S}_1 \rangle \right|^2.
\end{equation} 

In our implementation, a strong $\hat{S}_1$-polarized (vertical) coherent field serves as the LO, while the signal is encoded in the orthogonal polarization (horizontal) mode. The measurement of the Stokes operators $\hat{S}_2$ and $\hat{S}_3$ is performed using homodyne detection, which is formally equivalent to measuring two orthogonal quadratures, $\hat{X}$ and $\hat{P}$. To describe the signal generation and polarization transformation in our setup, we employ the Jones matrix formalism \cite{jones1941new}. The input field operator is prepared as vertically polarized state, 
\begin{equation}\label{eq0}
     \hat{E}_1 = 
     \hat{a}_{\mathrm{LO}}\begin{pmatrix}
         0 \\
         1
     \end{pmatrix},
\end{equation}
where $\hat{a}_{\mathrm{LO}}$ denotes the operator associated with the LO mode. After passing through an electro-optic amplitude modulator (AM), the field transforms as
\begin{equation}\label{eq1}
    \hat{E}_{2} = J_{\mathrm{AM}} \hat{E}_1,
\end{equation}
where the transfer matrix of the AM is given by
\begin{equation} \label{eq2}
     J_{\mathrm{AM}}= 
    \begin{pmatrix}
        \cos\frac{\phi}{2} & i\sin\frac{\phi}{2} \\
        i\sin\frac{\phi}{2} & \cos\frac{\phi}{2}
    \end{pmatrix},
\end{equation}
with $\phi = \frac{\pi V}{V_{\pi}}$ denoting the phase difference introduced between the polarization components, where $V_{\pi}$ is the half-wave voltage of the AM and $V$ is the applied voltage. Thus, the amount of modulation $\phi$ is controlled by $V$. From Equations.\eqref{eq1} and \eqref{eq2}, the field after the AM is given by
\begin{equation}\label{eq3}
    \hat{E}_2= \hat{a}_{\mathrm{LO}}
    \begin{pmatrix}
         i\,c_H \\
         c_V
    \end{pmatrix},
\end{equation}
where $c_H= \sin(\phi/2)$ and $c_V=\cos(\phi/2)$ represent the modulation coefficients associated with the $H$- and $V$-polarization modes, respectively. The AM thus introduces a horizontal component alongside the dominant vertical component, with its amplitude governed by $\phi$. For small applied voltages, $V \ll V_{\pi}$, $|c_V| \gg |c_H|$ ($\sim$40 dB), such that the $V$-polarized mode serves as a strong LO, while the weak $H$-polarized mode carries the encoded signal. 

A quarter-wave plate (QWP), with its fast axis at $0^\circ$, compensates the relative $\pi/2$ phase between the polarization components in Equation.\eqref{eq3}, resulting in an in-phase LO and signal. The corresponding output field transmitted to Bob is therefore given by
\begin{equation}\label{eq5}
    \hat{E}_3 = i\, \hat{a}_{\mathrm{LO}}
    \begin{pmatrix}
        c_H \\
        c_V
    \end{pmatrix}.
\end{equation}

At the receiver, the incoming field $\hat{E}_3$ passes through a sequence of waveplates arranged in a quarter($\lambda/4$)-half($\lambda/2$)-quarter($\lambda/4$) (QHQ) configuration, with fast axes oriented at $45^\circ$ (fixed), $\theta$ (variable), and $45^\circ$ (fixed) with respect to the horizontal, respectively. The corresponding transformation matrix of the QHQ configuration, obtained using Jones formalism, is given by 
\begin{equation}
    T_{\mathrm{QHQ}}= e^{i \pi/2}
    \begin{pmatrix}
        e^{-2i \theta}   & 0 \\
        0  &  -e^{2i \theta}
    \end{pmatrix},
\end{equation}
where $\theta$ determines the effective phase shift introduced between the orthogonal polarization modes. Applying this QHQ transformation to Equation.\eqref{eq5} yields
\begin{equation} \label{eq6}
\begin{gathered}
    \hat{E}_4 = -\, \hat{a}_{\mathrm{LO}}
    \begin{pmatrix}
       e^{-2i\theta}\,c_H \\
- e^{2i\theta}\,c_V
    \end{pmatrix} .
\end{gathered}
\end{equation}

The QHQ configuration thus enables control over the relative phase between the signal ($H$) and LO ($V$) components, given by $\Theta = 4\theta + \pi$. For $\theta = 0^\circ$, Equation.\eqref{eq6} reduces to
\begin{equation}\label{eq7}
    \hat{E}_4(0^\circ)= -\, \hat{a}_{\mathrm{LO}}
    \begin{pmatrix}
        c_H \\
        - c_V
    \end{pmatrix},
\end{equation}
which corresponds to a relative phase $\pi$ between the $H$ and $V$ components, therefore measuring the $X$-quadrature. For $\theta = 22.5^\circ$, Equation.\eqref{eq6} becomes
\begin{equation} \label{eq7.1}
    \hat{E}_4 (22.5^\circ)= \frac{\hat{a}_{\mathrm{LO}}}{\sqrt{2}}
    \begin{pmatrix}
        -(1-i)\, c_H\\
        (1+i)\, c_V
    \end{pmatrix},
\end{equation}
corresponding to a relative phase of $-\pi/2$, and hence enabling measurement of the $P$-quadrature. Notably, the additional $\pi$ term in $\Theta$ introduces a constant global phase offset, resulting in a relative phase of $\pi$ for $\theta=0^\circ$ and $-\pi/2$ for $\theta=22.5^\circ$. Since homodyne detection depends only on the relative phase modulo $\pi$, this offset does not affect the measured quadrature.

The interference between the signal and LO is realized using a half-wave plate (HWP) at $22.5^\circ$, followed by a polarizing beam splitter (PBS). Applying the HWP transformation to Equation.\eqref{eq7}, the field becomes
\begin{equation} \label{eq8}
    \hat{E}_5(0^\circ)= -\frac{\hat{a}_{\mathrm{LO}}}{\sqrt{2}}
    \begin{pmatrix}
        c_H - c_V \\
        c_H + c_V
    \end{pmatrix}.
\end{equation}
Similarly, for Equation.\eqref{eq7.1}, one obtains
\begin{equation} \label{eq9}
    \hat{E}_5(22.5^\circ) =
    \frac{\hat{a}_{\mathrm{LO}}}{2}
    \begin{pmatrix}
        (-1+i)\, c_H + (1+i)\,c_V \\
        (-1+i)\, c_H - (1+i)\,c_V
    \end{pmatrix}.
\end{equation}

The PBS separates the two orthogonal polarization components into distinct output ports, and the measurement corresponds to the difference in detected photon numbers
\begin{equation}\label{eq10}
\begin{aligned}
    \hat{n}_{\mathrm{out}} = \hat{a}_1^\dagger \hat{a}_1 - \hat{a}_2^\dagger \hat{a}_2 \\
    \langle \hat{n}_{\mathrm{out}} \rangle = |a_1|^2 - |a_2|^2,
\end{aligned}
\end{equation}
where $\hat{a}_1$ and $\hat{a}_2$ denote the output mode operators in the two paths. From Equation.\eqref{eq8}, 
\begin{equation}
    \hat{a}_1= -\frac{\hat{a}_{\mathrm{LO}}}{\sqrt{2}}(c_H - c_V), \quad
    \hat{a}_2=-\frac{\hat{a}_{\mathrm{LO}}}{\sqrt{2}}(c_H + c_V).
\end{equation}
Assuming a strong LO, $\hat{a}_{\mathrm{LO}}$ is approximated by the classical amplitude $\alpha_{\mathrm{LO}}$, which yields the $\hat{S}_2$($X$) measurement
\begin{equation}\label{eq11} 
    \langle \hat{S}_2 \rangle={\langle \hat{n}_{\mathrm{out}} \rangle}_{S_2} = -\, 2 \alpha_{\mathrm{LO}}^2 c_H c_V,
\end{equation}
substituting $c_H = \sin(\phi/2)$ and $c_V = \cos(\phi/2)$, this reduces to 
\begin{equation}\label{eq12} 
     \langle \hat{S}_2 \rangle = -\,\alpha_{\mathrm{LO}}^2 \sin \phi.
\end{equation}
For weak modulation ($V \ll V_\pi$, i.e., $\phi \ll 1$), 
\begin{equation}\label{eq13}
    \langle \hat{S}_2 \rangle \approx -\,\alpha_{\mathrm{LO}}^2 \phi \approx -\, \alpha_{\mathrm{LO}}^2 \frac{\pi V}{V_\pi},
\end{equation}
exhibiting a linear dependence of the measured signal on the applied voltage. Similarly, for the state in Equation.\eqref{eq9}, the corresponding $\hat{S}_3$($P$) measurement yields $\langle \hat{S}_3 \rangle = 0$, consistent with projection onto the conjugate quadrature. For intermediate values of $\theta$ (between $0^\circ$ and $22.5^\circ$), both $\hat{S}_2$ and $\hat{S}_3$ contribute. The normalized quadrature operators are therefore defined as
\begin{equation}\label{eq:p}
    \hat{X} = \frac{\hat{S}_2}{\sqrt{\langle \hat{S}_1 \rangle}}, \qquad
    \hat{P} = \frac{\hat{S}_3}{\sqrt{\langle \hat{S}_1 \rangle}},
\end{equation} 
where $\langle \hat{S_1} \rangle \approx \alpha_{\mathrm{LO}}^2$ represents the dominant LO intensity. For coherent states, the variances of $\hat{S}_2$ and $\hat{S}_3$ are equal. 

\subsection{Protocol Execution} \label{subsec:protocol}

We consider a GMCS UD-CVQKD protocol implemented in a prepare-and-measure configuration \cite{usenko2015unidimensional}, where only one quadrature is modulated while the orthogonal quadrature remains unmodulated. Without loss of generality, modulation is applied to the $X$-quadrature. For security analysis, the protocol follows an equivalent entanglement-based (EPR) representation \cite{grosshans2003virtual}. The implemented protocol proceeds as follows:
\begin{enumerate}
    \item Alice prepares a sequence of displaced coherent states $|\alpha_k\rangle$, where $\alpha_k = \frac{1}{\sqrt{2}}(x_{A,k} + i p_{A,k})$. The $X$ quadrature is Gaussian modulated as $x_{A,k}  \in \mathcal{N}(0, V_A)$, while $p_{A,k} = 0$. The total variance in the modulated quadrature is $V = V_A + 1$. 

    \item The prepared states are transmitted through an untrusted quantum channel characterized by transmittances $T_x$ and $T_p$, and channel excess noises $\xi_x$ and $\xi_p$ in the $X$ and $P$ quadratures, respectively, referred to the channel input. 

    \item Bob performs homodyne detection on the received states, characterized by detection efficiency $\eta$ and electronic noise $V_{el}$. The measured quadrature is selected via the LO phase, with $\phi_B = 0$ for the $X$ quadrature and $\phi_B = \pi/2$ for the $P$ quadrature, measured in separate acquisition runs.

    \item After quantum transmission, Alice and Bob publicly disclose a subset of their data to estimate the channel parameters $T_x$ and $\xi_x$. Since the $P$-quadrature is unmodulated, $T_p$ and $\xi_p$ cannot be directly estimated.

    \item Bob’s measurement outcomes in the $X$ quadrature, $x_B$, are continuous variables correlated with Alice’s data. These continuous outcomes are directly used for parameter estimation and subsequent classical post-processing, in accordance with the Gaussian-modulated CV-QKD framework.

    \item Finally, the secret key rate is evaluated using the estimated channel parameters and a reconciliation efficiency $\beta=0.95$.
\end{enumerate}

\subsection{Noise Models} \label{subsec:noisemodels}

In CV-QKD, system performance is determined by the channel transmittance and associated noise contributions. The total noise is decomposed into channel noise, $\chi^{\mathrm{ch}}$, referred to the channel input and potentially accessible to an eavesdropper, and detection noise, $\chi^{\mathrm{det}}$, arising from receiver imperfections and referred to the receiver input. All noise quantities are expressed in shot-noise units (SNU). The total noise, $\chi^{\mathrm{tot}}$, referred to the channel input, is given in each quadrature by \cite{scarani2009security}
\begin{equation}
    \chi_{x,p}^{\mathrm{tot}} = \chi_{x,p}^{\mathrm{ch}} + \frac{\chi^{\mathrm{det}}}{T_{x,p}}.
\end{equation}

\textit{Trusted detector model} (TD): In this model, the detection noise is assumed to be inaccessible to Eve and therefore does not contribute to her information. The channel excess noise depends only on the channel transmittance and noise sources external to Bob's detection system, and is given by \cite{scarani2009security, pan2025detector},
\begin{equation}
    \chi_{x,p}^{\mathrm{ch, TD}} = \frac{1 - T_{x,p}}{T_{x,p}} + \xi_{x,p},
\end{equation}
where $\frac{1 - T_{x,p}}{T_{x,p}}$ represents the vacuum noise introduced by channel loss, referred to the channel input. The excess noise $\xi_{x,p}$ accounts for contributions from both state preparation and the quantum channel. The detection noise on Bob’s side, determined by the detector efficiency $\eta$ and electronic noise $V_{el}$, is given by \cite{fossier2009improvement}
\begin{equation}
    \chi^{\mathrm{det, TD}}= \frac{(1-\eta) + V_{el}}{\eta}.
\end{equation}

\textit{Untrusted detector model} (UTD): In this model, the detector electronic noise is assumed to be accessible to Eve, while the detector inefficiency remains trusted. Consequently, the electronic noise of the detector is attributed to the channel, leading to a modified channel noise \cite{pan2025detector}
\begin{equation}
    \chi_{x,p}^{\mathrm{ch,UTD}} 
    = \frac{1 - T_{x,p}}{T_{x,p}} + \xi_{x,p} + \frac{V_{el}}{T_{x,p}\,\eta},
\end{equation}
with a reduced detection noise given by
\begin{equation}
    \chi^{\mathrm{det, UTD}}
    = \frac{1-\eta}{\eta}.
\end{equation}

In this work, the security performance is evaluated under both TD and UTD model assumptions to quantify the impact of detection noise on system performance. The TD model requires well-characterized and calibrated detectors, whereas the UTD model provides a conservative (worst-case) estimate of the secret key rate by attributing detector electronic noise to Eve. 

\section{Numerical Security Analysis} \label{sec:simulation}

The asymptotic lower bound secure key rate under Gaussian collective attacks and reverse reconciliation (RR) is given by \cite{usenko2015unidimensional, liao2018composable}
\begin{equation}\label{eq:k}
    K =\beta I_{\mathrm{AB}}-\chi_{\mathrm{BE}},
\end{equation}
where $\beta$ denotes the RR efficiency, $I_{\mathrm{AB}}$ is Shannon mutual information between Alice and Bob, and $\chi_{\mathrm{BE}}$ is the Holevo quantity that upper bounds Eve's accessible information \cite{holevo1999evaluating}. Under the assumption that Eve holds the purification of the joint state shared between Alice and Bob, the Holevo quantity becomes \cite{usenko2015unidimensional} 
\begin{equation}\label{eq:holevo}
    \chi_{\mathrm{BE}} = S({AB}) - S({A|x_B}),
\end{equation}
where $S(\cdot)$ denotes the von Neumann entropy. For Gaussian states, the entropy can be evaluated from the symplectic eigenvalues of the corresponding covariance matrices, as
\begin{equation} 
    \chi_{\mathrm{BE}} =
    G\!\left(\frac{\lambda_1 - 1}{2}\right) +
    G\!\left(\frac{\lambda_2 - 1}{2}\right) -
    G\!\left(\frac{\lambda_{\mathrm{cond}} - 1}{2}\right),
\end{equation}
where $G(x)= (x+1)\log_2(x+1) - x\log_2 x$, $\lambda_{1,2}$ are the symplectic eigenvalues of the covariance matrix $\gamma_{AB}$, and $\lambda_{\mathrm{cond}} = \sqrt{\det(\gamma_{A|x_B})}$ is the symplectic eigenvalue of the conditional covariance matrix $\gamma_{A|x_B}$. In the equivalent entanglement-based representation of the protocol, the covariance matrix after transmission through a lossy and noisy channel (characterized by $T_{x,p}, \xi_{x,p}$) is given by \cite{usenko2015unidimensional}
\begin{equation}
    \gamma_{AB} =
    \begin{pmatrix}
        \sqrt{1+V_A} & 0 & C_x & 0 \\
        0 & \sqrt{1+V_A} & 0 & C_p \\
        C_x & 0 & V_x^B & 0 \\
        0 & C_p & 0 & V_p^B
    \end{pmatrix},
\end{equation}
where the correlation term in the modulated $X$-quadrature is $C_x = \sqrt{T_x V_A}(1+V_A)^{1/4}$, and $C_p$ denotes the correlation in the unmodulated $P$-quadrature. Bob’s quadrature variances are given by
\begin{equation}
V_x^B = 1 + T_x (V_A + \xi_x), \quad
V_p^B = 1 + T_p\, \xi_p.
\end{equation} 

Following Bob’s homodyne measurement in the $X$-quadrature, the conditional covariance matrix is given by \cite{usenko2015unidimensional}
\begin{equation}
    \gamma_{A|x_B} =
    \begin{pmatrix}
        \frac{\sqrt{1+V_A}(1 + T_x \xi_x)}{1 + T_x (V_A + \xi_x)} & 0 \\
        0 & \sqrt{1+V_A}
    \end{pmatrix}.
\end{equation}

Since the $P$-quadrature is not modulated in the unidimensional protocol, the correlation parameter $C_p$ cannot be directly estimated from experimental data. Therefore, in the numerical analysis, $\chi_{\mathrm{BE}}$ is evaluated under a worst-case assumption by maximizing over all physically admissible values of $C_p$, i.e., $\chi_{\mathrm{BE}}^{\mathrm{worst}}=\max_{C_p} (\chi_{\mathrm{BE}})$, while $V_p^B$ is directly measured at the receiver.  

The Shannon mutual information between Alice and Bob is expressed as
\begin{equation}\label{eq:MI}
    I_{\mathrm{AB}}=\frac{1}{2}\log_2 \left(1 + \frac{T_x \eta V_A}{1 + T_x \eta \,\xi_x +V_{el}}\right),
\end{equation}
which depends only on the total noise $\chi^{\mathrm{tot}}$ affecting Bob’s measurement and not on how this is distributed between $\chi^{\mathrm{ch}}$ and $\chi^{\mathrm{det}}$. Consequently, $I_{\mathrm{AB}}$ remains unchanged under both TD and UTD models, whereas $\chi_{\mathrm{BE}}$ explicitly depends on the adopted noise model.

According to the Heisenberg uncertainty principle, the parameters $V_p^B$ and $C_p$ are constrained by the physicality condition of the covariance matrix, which must satisfy \cite{weedbrook2012gaussian}
\begin{equation} \label{eq14}
    \gamma_{AB} + i\Omega \geq 0,
\end{equation}
where $\Omega = \bigoplus_{i=1}^{n} \omega, \quad
\omega =
\begin{pmatrix}
0 & 1 \\
-1 & 0
\end{pmatrix}$.
This condition ensures that the covariance matrix corresponds to a physically realizable quantum state. 

\begin{figure*}[tb]
    \centering
    \includegraphics[scale = 0.62]{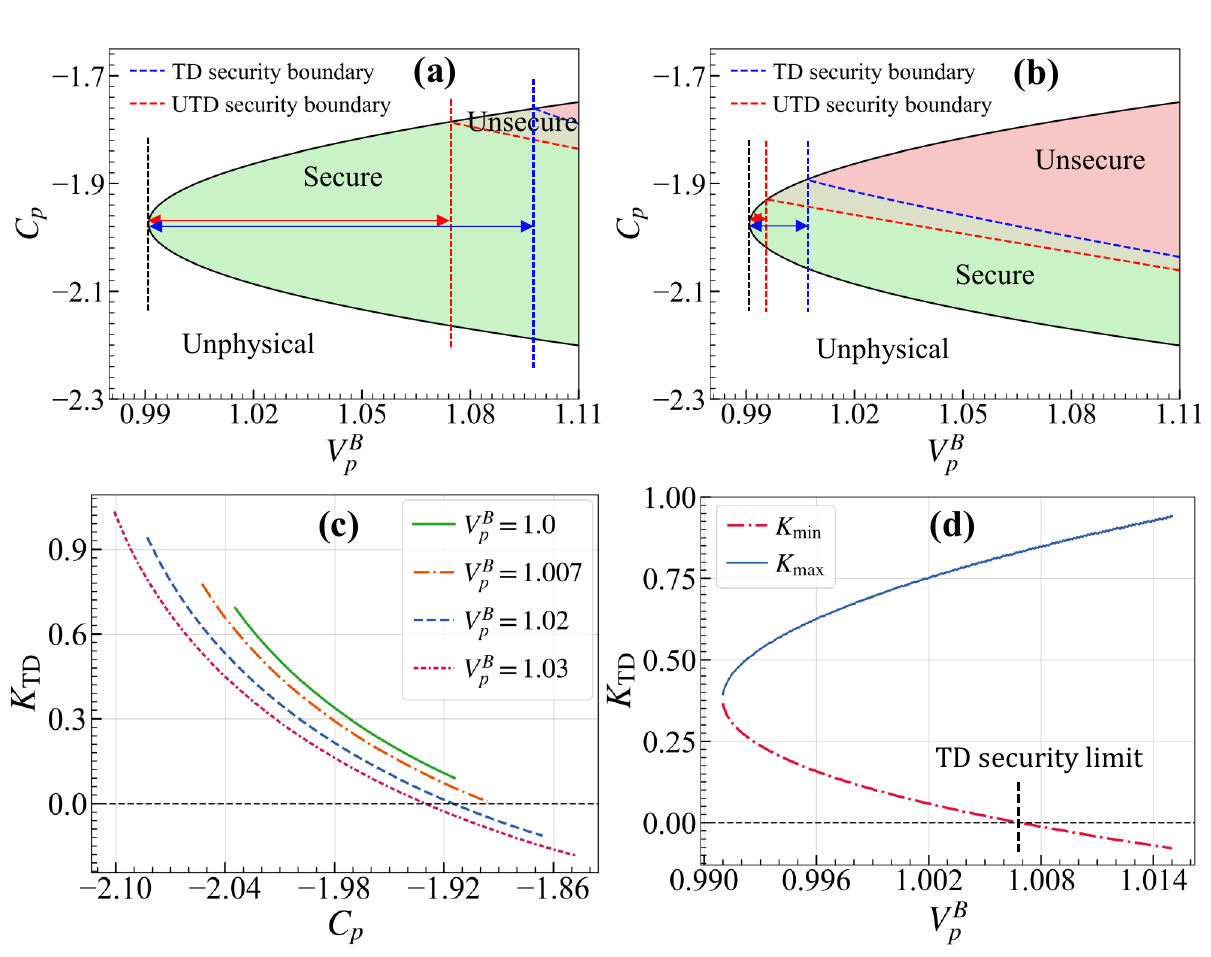}
    \caption{(a) Physical and secure regions in the $(V_p^B, C_p)$ parameter space for both TD and UTD models. The solid black curve denotes the physicality boundary. The red and blue dashed curves represent the zero key-rate ($K=0$) boundaries for the UTD and TD cases, respectively, separating secure ($K>0$) and unsecure ($K<0$) regions, shown in green and red shades. The region between the vertical dotted lines indicates the $C_p$-independent security threshold for both models. Results are shown for a low electronic-noise detector with $V_{el} = 0.1$ SNU.
    (b) Same as in (a), but for a high electronic-noise detector with $V_{el} = 1.4$ SNU. For higher $V_{el}$, this $C_p$-independent secure region shrinks significantly in the UTD case, while remaining comparatively robust for the TD model.
    (c) Secret key rate for the TD model, $K_{\mathrm{TD}}$, as a function of the unknown parameter $C_p$, for representative values of $V_p^B$.
    (d) Extremal TD key rates, $K_{\min}$ and $K_{\max}$, corresponding to pessimistic and optimistic values of $C_p$, respectively, as functions of $V_p^B$.
    In all panels, the system parameters are $V_A = 20$ SNU, $T_x = 0.91$, $\xi_x = 0.01$ SNU, and $\eta = 0.81$.}
    \label{fig:security}
\end{figure*}

Figures~\ref{fig:security}(a) and \ref{fig:security}(b) illustrate the physical and secure regions in the $(V_p^B, C_p)$ parameter space for low ($V_{el}=0.1$ SNU) and high ($V_{el}=1.4$ SNU) electronic-noise detectors, respectively, under both TD (blue) and UTD (red) models. Here, $V_{el}=0.1$ corresponds to a 10 dB shot-noise clearance, while $V_{el}=1.4$ corresponds to an experimentally measured shot-noise clearance of approximately 2.4 dB. The remaining parameters, $V_A=20$ SNU, $T_x=0.91$, $\xi_x=0.01$ SNU, and $\eta=0.81$, are chosen according to the experimental operating conditions. The boundary imposed by the physicality condition (Equation.\eqref{eq14}) exhibits a parabolic shape with vertex $(V_0^B, C_0) =(0.99, -1.97)$, determined by the total system noise. The parameter space is divided into unphysical, physical but unsecure ($K < 0$), and secure ($K > 0$) regions. Since $C_p$ remains experimentally inaccessible, a conservative security region is identified by considering the range of $V_p^B$ for which the key rate remains positive for all physically admissible values of $C_p$. This corresponds to the interval between the vertex (vertical black dashed line) and the model-dependent thresholds (red and blue dashed lines for UTD and TD models, respectively). Within this $V_p^B$ interval, the protocol guarantees secure key generation irrespective of $C_p$ values. Beyond these thresholds, security depends on the specific value of $C_p$, and the worst-case assumption no longer ensures a positive key rate. For the low-noise detector (Figure~\ref{fig:security}(a)), the secure region extends up to $V_p^B \approx 1.075$ (UTD) and $V_p^B \approx 1.098$ (TD), providing a comparatively broad operational regime. As $V_{el}$ increases, this $C_p$-independent secure region shrinks significantly. For the $V_{el}=1.4$ case (Figure~\ref{fig:security}(b)), the reduction is particularly pronounced for the UTD model, where the secure region is restricted to $V_p^B \leq 0.996$. The TD model, however, retains a comparatively wider operational regime, $V_p^B \leq 1.007$, even under high detector electronic-noise constraints.

Since the secure region in the UTD model becomes extremely narrow ($V_p^B \leq 0.996$), whereas physically $V_p^B \geq 1$, no physically accessible $C_p$-independent secure regime exists under high $V_{el}$ conditions. The subsequent analysis therefore focuses on the TD model. Figure~\ref{fig:security}(c) shows the TD secret key rate, $K_{\mathrm{TD}}$, as a function of $C_p$ for representative values of $V_p^B$. For each $V_p^B$, the key rate reaches a maximum at an optimal correlation and decreases with increasing $C_p$. Within the security-threshold regime, $V_p^B \leq 1.007$, $K_{\mathrm{TD}}$ remains positive over the entire physically allowed range of $C_p$, ensuring security without explicit estimation of $C_p$. For $V_p^B > 1.007$, $K_{\mathrm{TD}}$ becomes negative over part of the physical $C_p$ range, implying that security cannot be guaranteed under worst-case assumptions. Although larger $V_p^B$ can produce higher key rates for certain $C_p$, $V_p^B$ values closer to unity provide more robust performance under worst-case conditions. Figure~\ref{fig:security}(d) presents the extremal key rates, $K_{\min}=\min_{C_p} K(V_p^B, C_p)$ and $K_{\max}= \max_{C_p} K(V_p^B, C_p)$, as functions of $V_p^B$, obtained by optimizing over $C_p$ within the physical region. The increasing separation between $K_{\max}$ and $K_{\min}$ with $V_p^B$ indicates a greater sensitivity of the protocol security to the unknown correlation parameter $C_p$.

\begin{figure}[h]
    \centering
    \includegraphics[scale = 0.43]{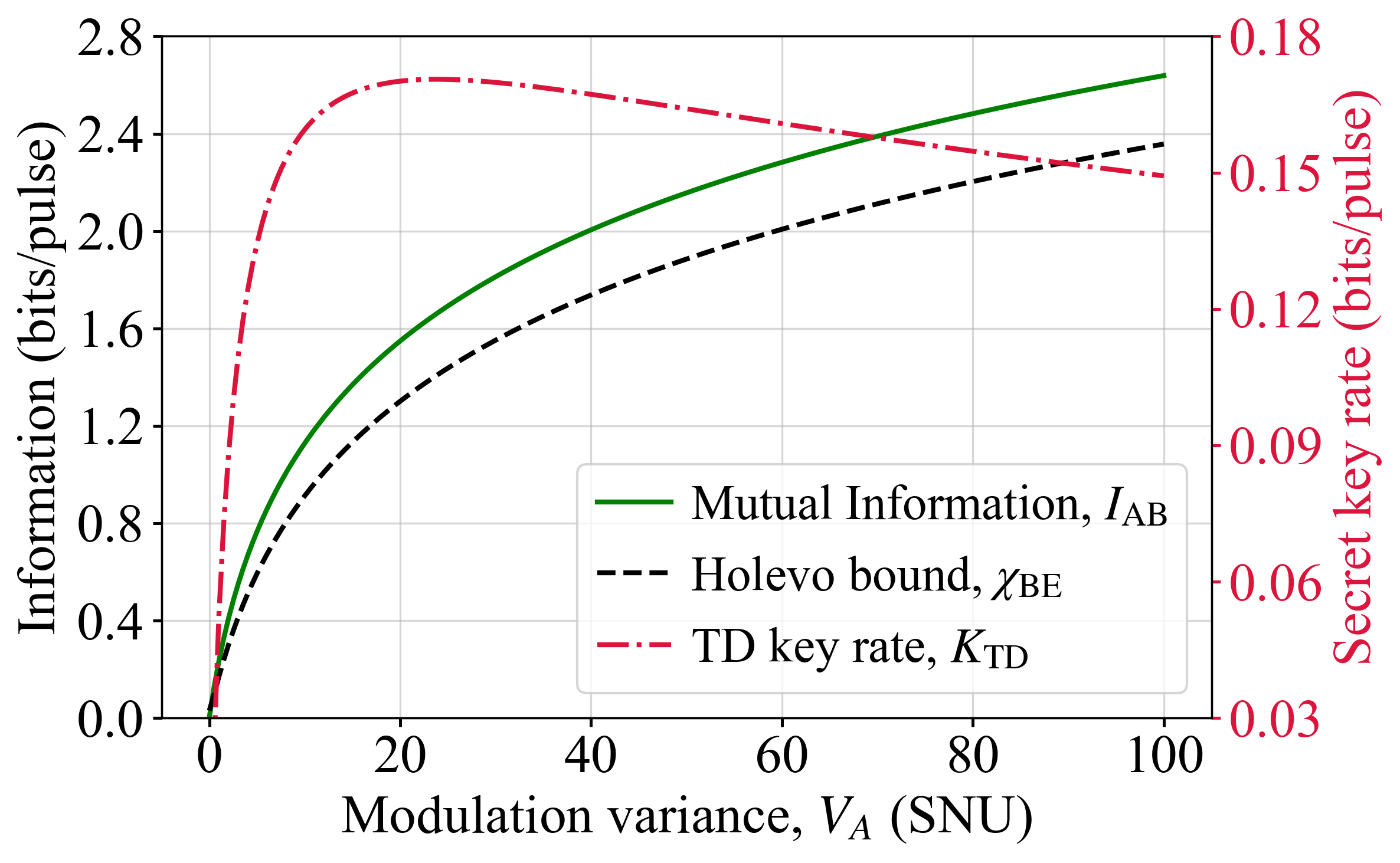}
    \caption{Behavior of the mutual information $I_{\mathrm{AB}}$, Holevo bound $\chi_{\mathrm{BE}}$, and TD secret key rate $K_{\mathrm{TD}}$ as functions of the modulation variance $V_A$. The parameters are $V_p^B=1.003$, $T_x = 0.91$, $\xi_x = 0.01$ SNU, $\eta = 0.81$, and $V_{el} = 1.4$ SNU, assuming worst-case $C_p$.}
    \label{fig:K-VA}
\end{figure}

We numerically evaluate the performance of the protocol in terms of $I_{\mathrm{AB}}$, $\chi_{\mathrm{BE}}$, and $K_\mathrm{TD}$ as functions of the modulation variance $V_A$ using Equations.\eqref{eq:MI}, \eqref{eq:holevo}, and \eqref{eq:k}, respectively, as shown in Figure~\ref{fig:K-VA}. The mutual information increases with $V_A$ due to improved SNR, with a sublinear trend at large $V_A$ due to its logarithmic dependence. The Holevo bound also increases, reflecting enhanced information leakage to Eve. Consequently, $K_\mathrm{TD}$ initially increases, reaches a maximum at $V_A^{\mathrm{opt}} \approx 23.63$ SNU with $K_{\max} \approx 0.17$ bits/pulse, and decreases thereafter as $\chi_{\mathrm{BE}}$ approaches and eventually exceeds $\beta I_{\mathrm{AB}}$ (for $\beta < 1$). Notably, the secret key rate remains within 99\% of its maximum over a broad range $V_A \in [18.60, 30.66]$ SNU, indicating robustness to modulation settings. This behavior highlights the importance of optimizing $V_A$ for practical implementations. The $V_A^{\mathrm{opt}}$ depends on channel and detection parameters and may shift with variations in loss, excess noise, and RR efficiency.

\begin{figure*}[tb]
    \centering
    \includegraphics[scale = 0.39]{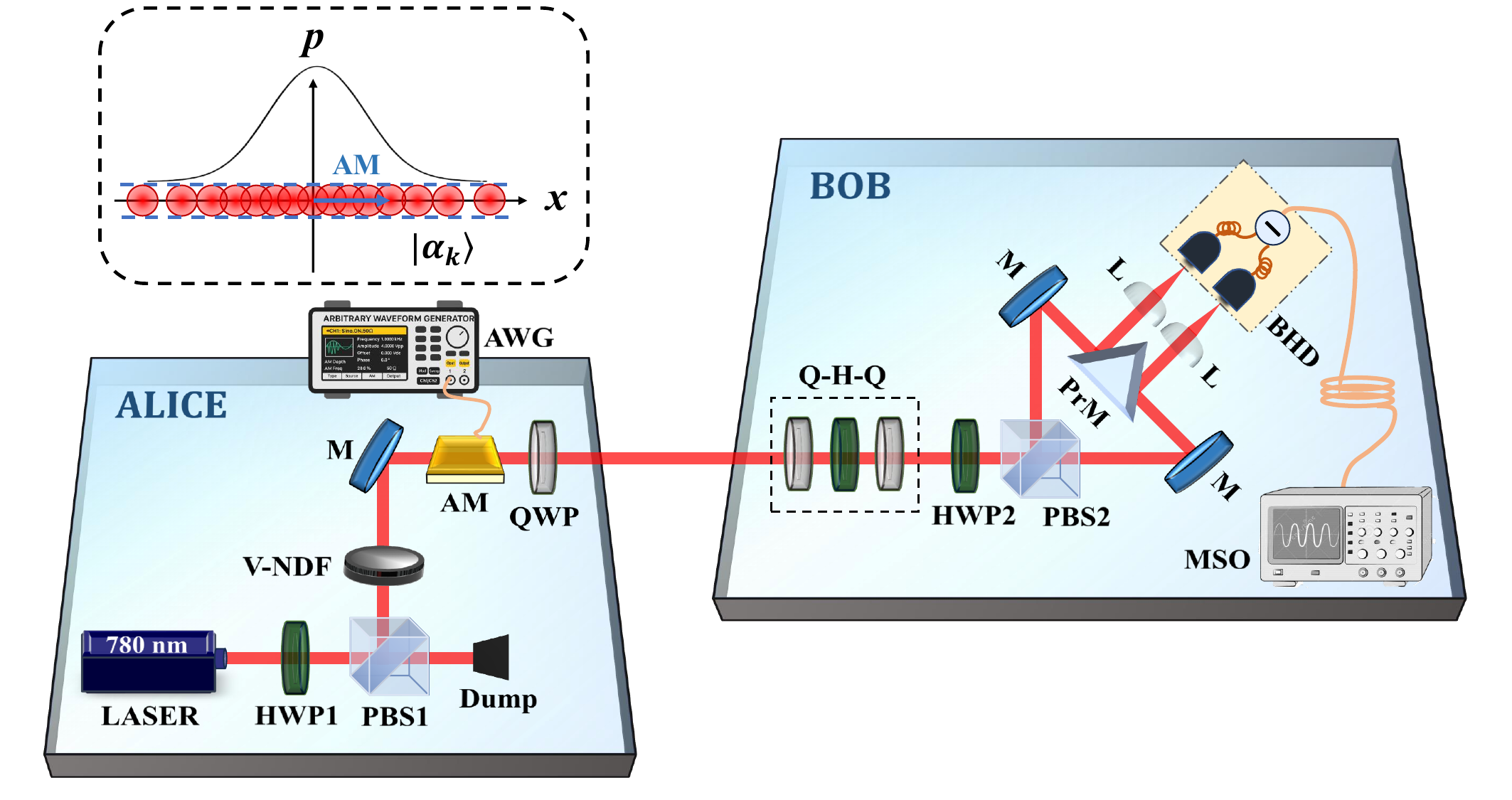}
    \caption{Schematic of the experimental setup for Gaussian-modulated UD-CVQKD system over a free-space channel. HWP: half-wave plate; PBS: polarizing beam-splitter; V-NDF: variable neutral density filter; M: mirror; AM: amplitude modulator; QWP: quarter-wave plate; AWG: arbitrary waveform generator; Q-H-Q: QWP-HWP-QWP configuration; PrM: prism mirror; L: lens; BHD: balanced homodyne detector; MSO: mixed signal oscilloscope. Inset: phase-space representation of unidimensional Gaussian modulation in the $X$-quadrature.}
    \label{fig:expt}
\end{figure*}

\section{Experimental setup}\label{sec:experimentalsetup}
We employed a laboratory free-space as the quantum channel for UD-CVQKD implementation, modulated with the Gaussian distribution. Figure~\ref{fig:expt} presents the schematic of the experimental setup used to implement the protocol described in Section~\ref{subsec:protocol}, including the configurations for Alice and Bob.

\subsection{Alice}
A continuous-wave laser (Sacher Lasertechnik) operating at 780 nm serves as the coherent source at Alice. The beam is spatially mode-cleaned to produce a Gaussian mode suitable for free-space propagation. A vertically polarized field is prepared using HWP1 and PBS1, with its power controlled by a variable neutral density filter (V-NDF). Gaussian modulation is implemented using an electro-optic AM (Thorlabs EO-AM-NR-C1), driven by an arbitrary waveform generator (AWG, Tektronix AWG5204) operating at a repetition frequency of 2.5 MHz with a sampling rate of 2.5 GSa/s. A low-amplitude square pulse train (few mV, 50\% duty cycle) is applied to the AM, with each pulse containing 1000 samples. The modulation pulse width is 200 ns, followed by an equal-duration zero-voltage interval serving as a vacuum reference. As described in Equations.\eqref{eq0}-\eqref{eq3}, the applied modulation induces a weak horizontally polarized component ($\ket{H}$) carrying the Gaussian-modulated signal, while the vertically polarized LO ($\ket{V}$) remains dominant. The modulation depth (1 to 2000 mV) is kept well below the half-wave voltage of the AM ($V_\pi = 260$ V), ensuring linear operation (Equation.\eqref{eq13}). The QWP compensates for the additional $\pi/2$ phase shift between the signal and the LO. The modulation voltage is generated from a computer-controlled Gaussian random sequence with variance $V_A$. The resulting Gaussian-modulated polarized coherent states, \( |\alpha_k\rangle \), are then transmitted through the free-space channel to the receiver (Bob). 

\subsection{Bob}
At the receiver, the incoming states are measured using balanced homodyne detection implemented in the Stokes-parameter basis. The measurement basis is selected using a QWP-HWP-QWP (Q-H-Q) configuration, which provides a controllable relative phase between the LO and the signal. This enables selection between measurements of the Stokes operators $\hat{S}_2$ and $\hat{S}_3$, corresponding to the $\hat{X}$ and $\hat{P}$ quadratures, respectively. Specifically, when both QWPs are oriented at $45^\circ$ and the HWP is set at $0^\circ$, the system measures $\hat{S}_2$ ($\hat{X}$). Rotating the HWP to $22.5^\circ$ enables the measurement of $\hat{S}_3$ ($\hat{P}$), as described in Equations.\eqref{eq7} and \eqref{eq7.1}. Thus, the Q-H-Q configuration functions as a basis switcher between the $\hat{X}$ and $\hat{P}$ quadratures. HWP2, set at $+22.5^\circ$, rotates the polarization basis such that the vertically polarized LO ($\hat{S}_1$) and the weak horizontally polarized signal interfere in the diagonal/anti-diagonal or circular bases. The fields are then analyzed using PBS2, which separates the orthogonal polarization components and directs them to a balanced homodyne detector (BHD, Thorlabs PDB435A-AC) via mirrors (M), a prism mirror (PrM), and lenses (L). The photodiodes have a responsivity of 0.51 A/W at 780 nm, corresponding to a detection efficiency of approximately 0.81 per diode. The difference photocurrent from the BHD (Equation.\eqref{eq13}) is recorded using a mixed-signal oscilloscope (MSO, Tektronix MSO68B) and transferred to a computer for post-processing. 

\subsection{Data acquisition}
The experiment was performed using different Gaussian random sequences, with the modulation variance varied over the range $V_A \in [4, 97]$ SNU. For each $V_A$ setting, approximately $2 \times 10^5$ valid optical pulses were recorded within a single acquisition window of 100 ms. For each pulse event, the central 80\% of the recorded samples (excluding 10\% at each edge) were integrated over the pulse duration to obtain a single quadrature value per pulse, corresponding to Bob’s measurement outcome, $x_B$. Following data acquisition, the recorded data were sorted according to the measured quadrature basis to obtain correlated continuous-variable data sets. The $\hat{X}$-quadrature measurements were used for secure key generation, whereas the $\hat{P}$-quadrature data were used for security verification. 

\section{Results and Discussions} \label{sec:rnd}

The detector was first characterized in terms of detection noise (electronic noise and detection efficiency) and shot-noise calibration. The shot-noise level was determined by measuring the variance of the BHD output by interfering the LO with vacuum (AM off), exhibiting a linear dependence on the LO power used for calibration. The operational LO power ($V$-polarization) was then fixed at $P_{\mathrm{LO}}$ for all subsequent measurements, ensuring operation within the linear response regime of the detector. At this operating point, the shot-noise variance, obtained after subtracting the electronic noise contribution from the total measured variance, was $N_0 = 2.905~\mathrm{mV}^2$, which is defined as 1 SNU. The electronic noise variance, measured with both the signal and LO blocked, was $V_{el} = 4.058~\mathrm{mV}^2$, corresponding to $V_{el} = 1.397$~SNU after normalization. Due to detector saturation at higher LO powers, the accessible LO range was limited, resulting in operation in an electronic-noise-dominated regime ($V_{el} > N_0$). Nevertheless, a clearance of $\sim$2.4 dB between the total shot-noise variance (including electronic noise) and the electronic noise floor was achieved, ensuring resolvable detection above $V_{el}$. 

\begin{figure}[h]
    \centering
    \includegraphics[scale = 0.45]{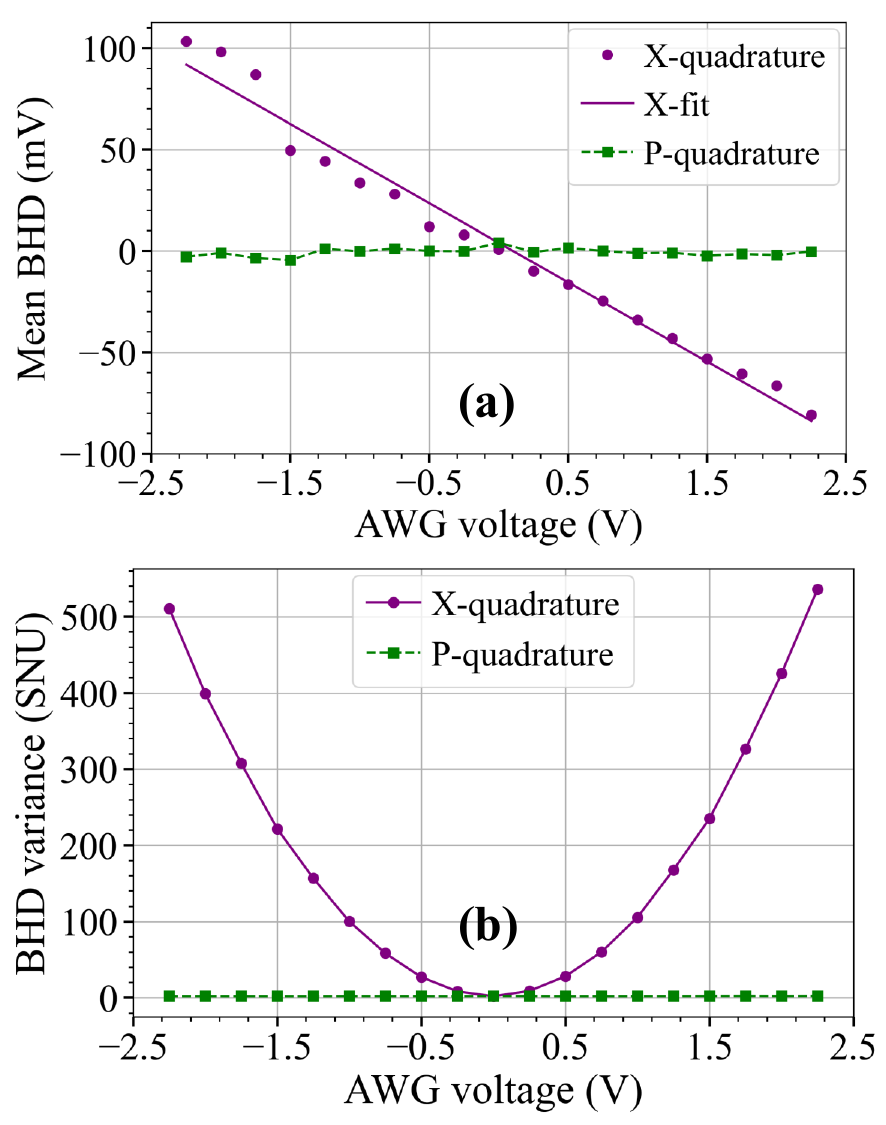}
    \caption{Calibration of the amplitude modulation and detection system. (a) Mean BHD output as a function of the applied AWG voltage (-2.25 V to +2.25 V), exhibiting a linear response for the modulated $X$-quadrature, while the $P$-quadrature remains approximately constant. (b) Corresponding variance with AWG voltage, showing a quadratic dependence for $X$ and nearly constant response for $P$.}
    \label{fig:am}
\end{figure}

\begin{figure}[h]
    \centering
    \includegraphics[scale = 0.45]{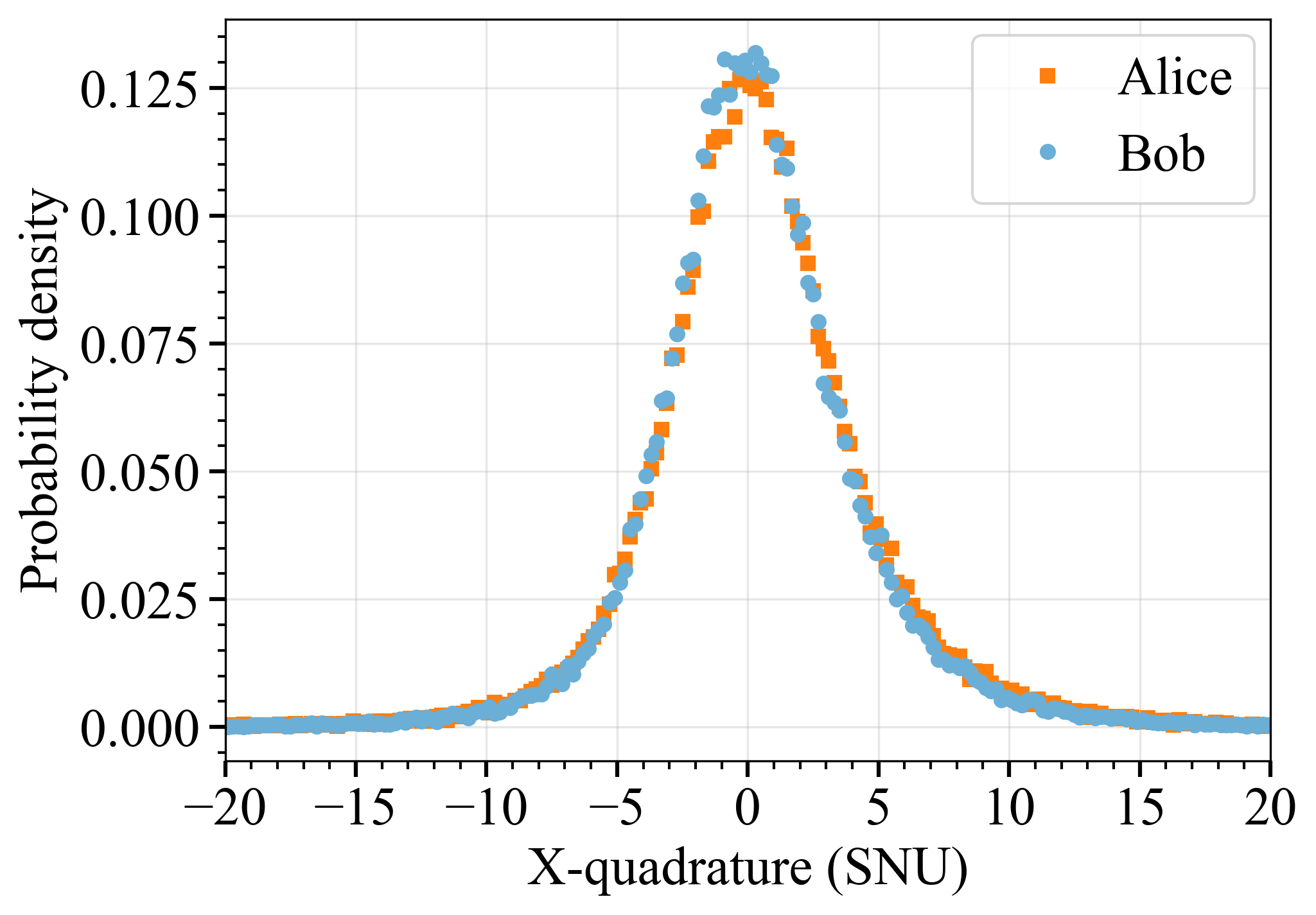}
    \caption{Quadrature probability distributions for Alice and Bob at $V_A=17.54$ SNU. Alice’s distribution is Gaussian with variance set by the applied modulation, while Bob’s measured distribution shows reduced variance ($V_x^B=16.73$ SNU) due to channel loss.}
    \label{fig:bob}
\end{figure}

Following detector calibration, the transmitter-side modulation was characterized. To establish a quantitative relation between the applied modulation and the corresponding measured quadrature at Bob, the system linearity was verified in the small-modulation regime ($V \ll V_\pi$), consistent with Equation.\eqref{eq13}. For calibration, a set of known voltage levels was applied to the AM, and the corresponding BHD outputs were recorded over approximately $2\times10^4$ pulses. As shown in Figure~\ref{fig:am}(a), the mean BHD response exhibits a linear dependence on the applied AWG voltage for the modulated $X$ quadrature, while the unmodulated $P$ quadrature remains nearly constant within experimental fluctuations, confirming unidimensional modulation and linear system response. A linear fit to the $X$-quadrature data yields a conversion factor $k = -3.6453 \times 10^{-2}$, enabling mapping of the applied modulation voltage to quadrature units and estimation of the modulation variance $V_A$ in SNU. In addition, the variance of the $X$-quadrature shows a quadratic dependence on the applied voltage, as shown in Figure~\ref{fig:am}(b).

\begin{figure}[h]
    \centering
    \includegraphics[scale = 0.48]{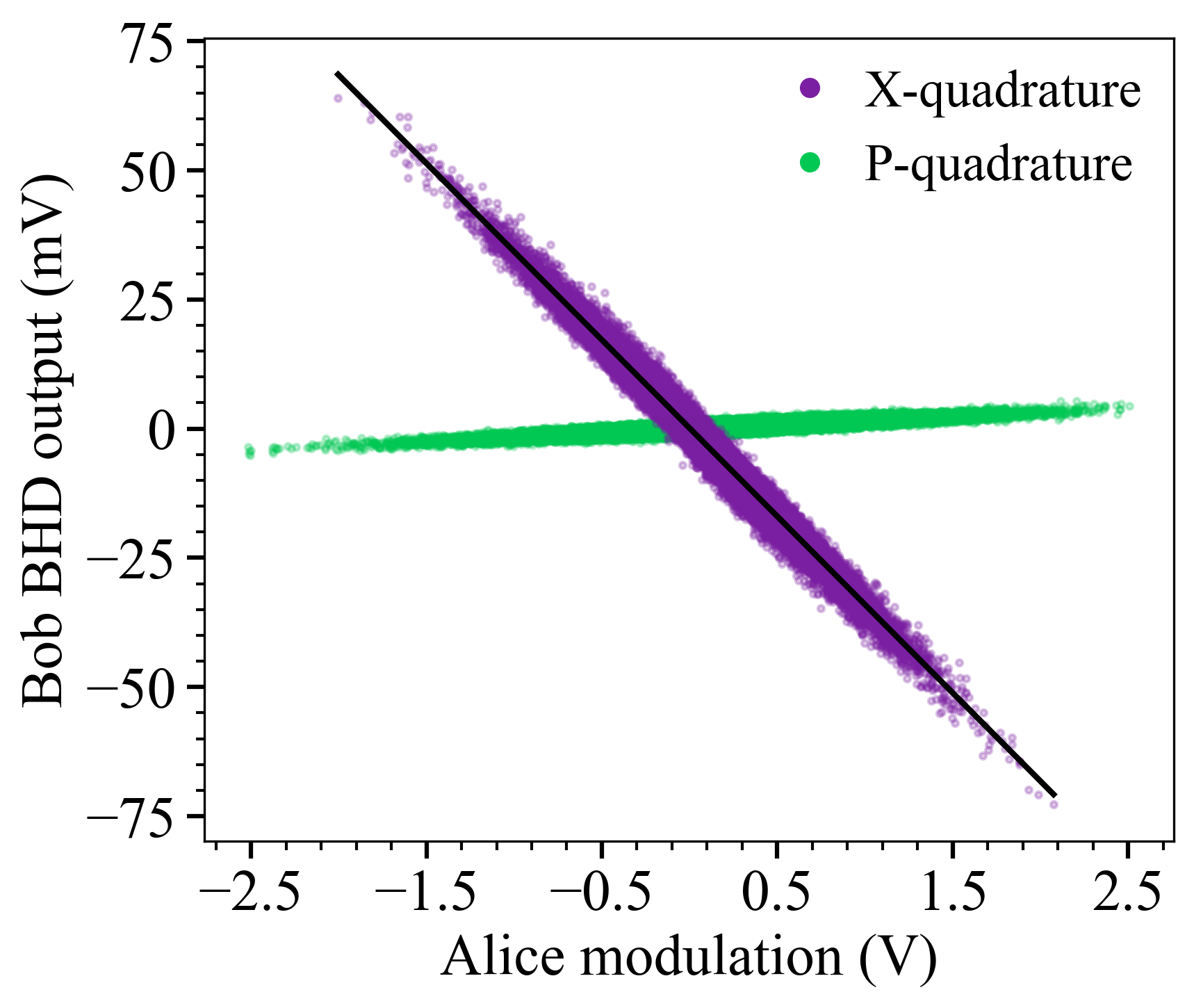}
    \caption{Correlation between Alice’s modulation voltage and Bob’s measured output. A clear linear relationship is observed for the modulated $X$-quadrature, with a correlation coefficient of $0.98$, while the $P$-quadrature remains nearly constant.}
    \label{fig:scatter}
\end{figure}

For protocol implementation, Gaussian-distributed modulation with zero mean and varying variance $V_A$ was applied to encode the signal states. The modulation variance was varied in the range $0.01$ to $0.175~\mathrm{V}^2$, corresponding to approximately $4.01$ to $96.88$ SNU. The total system transmittance, including channel and optical component losses, was measured to be $0.91$ using a power meter. The Gaussian nature of the prepared and measured states was verified from the quadrature distributions. As shown in Figure~\ref{fig:bob}, both Alice’s and Bob’s data exhibit Gaussian profiles for $V_A=17.54$ SNU, with Bob’s distribution showing reduced variance ($V_x^B=16.73$ SNU) due to channel loss. The correlation between Alice’s modulation and Bob’s measurement outcomes is presented in Figure~\ref{fig:scatter}. The modulated $X$-quadrature exhibits a clear linear correlation, while the unmodulated $P$-quadrature remains narrowly distributed. A linear regression of the $X$-quadrature data yields a correlation coefficient of approximately 0.98, confirming stable signal transfer.  

The channel parameters $T_x$ and $\xi_x$ were estimated using the full dataset of approximately $2 \times 10^{5}$ pulses, ensuring statistical reliability. These parameters were obtained from the covariance relations between Alice’s and Bob’s quadrature data. The channel transmittance is calculated as \cite{shen2019free}
\begin{equation}\label{eq:tr}
T_x = \frac{\mathrm{Cov}(x_A, x_B)^2}{V_A^2},
\end{equation}
where $x_A$ and $x_B$ denote Alice’s and Bob’s quadrature values (in SNU), respectively. Under the TD and UTD models, the excess noise is estimated as
\begin{equation}\label{eq:xi}
\xi_x^{\mathrm{TD}} = \frac{\mathrm{Var}(x_B) - V_{el} - 1}{\eta T_x} - V_A,
\end{equation}
\begin{equation}
\xi_x^{\mathrm{UTD}} = \frac{\mathrm{Var}(x_B) - 1}{T_x} - V_A.
\end{equation}
The estimated average channel transmittance, over varying $V_A$, is $T_x \approx 0.882$, in good agreement with the value obtained from power meter measurements ($0.91$). The excess noise remains nearly constant, with an average value of $\xi_x^{\mathrm{TD}} \approx 0.0105$ SNU under the TD model. In contrast, under the UTD model, the excess noise increases significantly ($\xi_x^{\mathrm{UTD}} > 1$ SNU), as the detector electronic noise is attributed to a potential eavesdropper.

\begin{figure}[h]
    \centering
    \includegraphics[scale = 0.48]{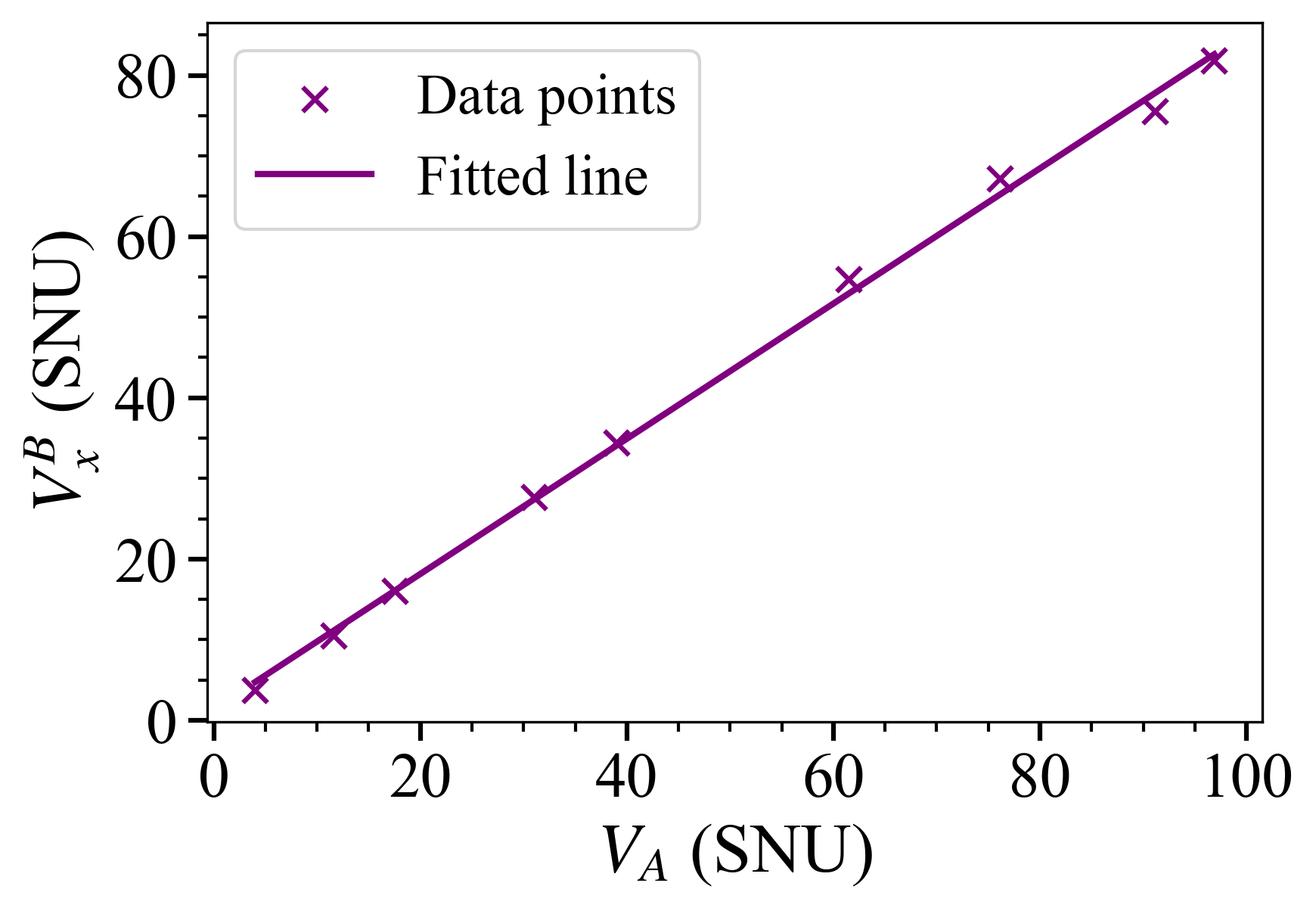}
    \caption{Bob’s measured $X$-quadrature variance $V_x^B$ as a function of Alice’s modulation variance $V_A$, with all variances expressed in SNU. The experimental data (markers) are fitted with a linear model (solid line), showing a linear dependence of $V_x^B$ on $V_A$.}
    \label{fig:correlation}
\end{figure}

The measured variance of Bob’s $X$-quadrature, $V_x^B$, is plotted as a function of $V_A$ (both in SNU) in Figure~\ref{fig:correlation}, showing a linear dependence consistent with Equation.\eqref{eq:xi}. The variance of the $P$-quadrature at the receiver is obtained from the  $S_3$ measurement (Equation.\eqref{eq:p}), yielding $V_p^B=1.003$ SNU. The experimentally determined critical TD limit is found to be $({V_p^B})_{\max}=1.0064$, in close agreement with the theoretical limit of $1.007$. Using the experimentally estimated parameters, the secure key rate is evaluated for different values of $V_A$. Under the UTD model, no positive secure key rate is obtained in the high electronic-noise regime of the detector. This is due to the significantly increased excess noise $\xi_x^{\mathrm{UTD}}$, which restricts the $C_p$-independent secure region to ${V_p^B} \le 0.996$ (theoretically) in Figure~\ref{fig:security}(b). Since this condition is not satisfied experimentally, no secret key is achievable under the worst-case $C_p$ assumption for any value of $V_A$. The TD model, however, allows for positive secure key generation over a wide range of modulation variances up to $V_A = 76.11$ SNU. In the following, we therefore restrict our analysis and discussion to the TD model. 

Since the key rate strongly depends on the unknown correlation parameter $C_p$ (see Figure~\ref{fig:security}(c)), a worst-case analysis is performed by minimizing over all physically allowed values of $C_p$ when evaluating the Holevo bound $\chi_{\mathrm{BE}}$ and the corresponding secret key rate. The extracted parameters, including $T_x$, $\xi_x$, $I_{\mathrm{AB}}$, $\chi_{\mathrm{BE}}$, and the resulting worst-case key rate $K_{\mathrm{TD}}$, are summarized in Table~\ref{tab:udcvqkd_params}. The transmittance, calculated using Equation.\eqref{eq:tr}, varies within $T_x \in [0.865,\,0.898]$ across different values of $V_A$, while the corresponding excess noise lies within $\xi_x \in [0.0096,\,0.0124]$ SNU. The TD secret key rate initially increases with $V_A$, reaches a maximum, and subsequently decreases, becoming negative at higher modulation values. This behavior is primarily associated with a reduction in the estimated transmittance $T_x$ (to $0.815$ and $0.832$) at large $V_A$ ($91.10$ and $96.88$ SNU), while the corresponding excess noise $\xi_x$ remains nearly unchanged ($0.00965$ and $0.00994$) SNU; these points are not included in the table. The observed reduction in $T_x$ with increasing $V_A$ can be attributed to the sub-linear scaling of the measured covariance (Equation.\eqref{eq:tr}) due to detector non-idealities and residual technical noise at larger modulation depths, leading to an underestimation of the normalized covariance used for transmittance estimation. Although $I_{\mathrm{AB}}$ remains slightly higher than $\chi_{\mathrm{BE}}$, the finite RR efficiency, $\beta = 0.95$, results in $\beta I_{\mathrm{AB}} < \chi_{\mathrm{BE}}$, leading to negative $K_{\mathrm{TD}}$ at high modulation. 

\begin{table}[h]
    \centering
    \caption{Experimentally estimated security parameters for different modulation variances ($V_A$) under the TD assumption. $V_A$ and $\xi_x$ are presented in SNU, while the information quantities ($I_{\mathrm{AB}}$, $\chi_{\mathrm{BE}}$) and the secret key rate ($K_{\mathrm{TD}}$) are expressed in bits per pulse. The reported key rate corresponds to the worst-case estimate over $C_p$.}
    \renewcommand{\arraystretch}{1.3}
    \setlength{\tabcolsep}{5pt} 
    \label{tab:udcvqkd_params}
    \begin{tabular}{c c c c c c c}
        \hline\hline
        $V_A$ & $T_x$ & $\xi_x$ & SNR & $I_{\mathrm{AB}}$ & $\chi_{\mathrm{BE}}$ & $K_{\mathrm{TD}}$ \\
        \hline
        4.01  & 0.898 & 0.0106 & 1.496  & 0.6597 & 0.5260 & 0.1007 \\
        11.57 & 0.891 & 0.0111 & 4.284  & 1.2009 & 1.0326 & 0.1082 \\
        17.54 & 0.896 & 0.0124 & 6.525  & 1.4558 & 1.2863 & 0.0968 \\
        30.98 & 0.878 & 0.0099 & 11.309 & 1.8108 & 1.6481 & 0.0722 \\
        38.99 & 0.865 & 0.0096 & 14.024 & 1.9546 & 1.8220 & 0.0348 \\
        61.47 & 0.876 & 0.0102 & 22.379 & 2.2736 & 2.1188 & 0.0411 \\
        76.11 & 0.869 & 0.0096 & 27.501 & 2.4165 & 2.2681 & 0.0275 \\
        \hline\hline
    \end{tabular}
\end{table}

\begin{figure}[h]
    \centering
    \includegraphics[scale = 0.41]{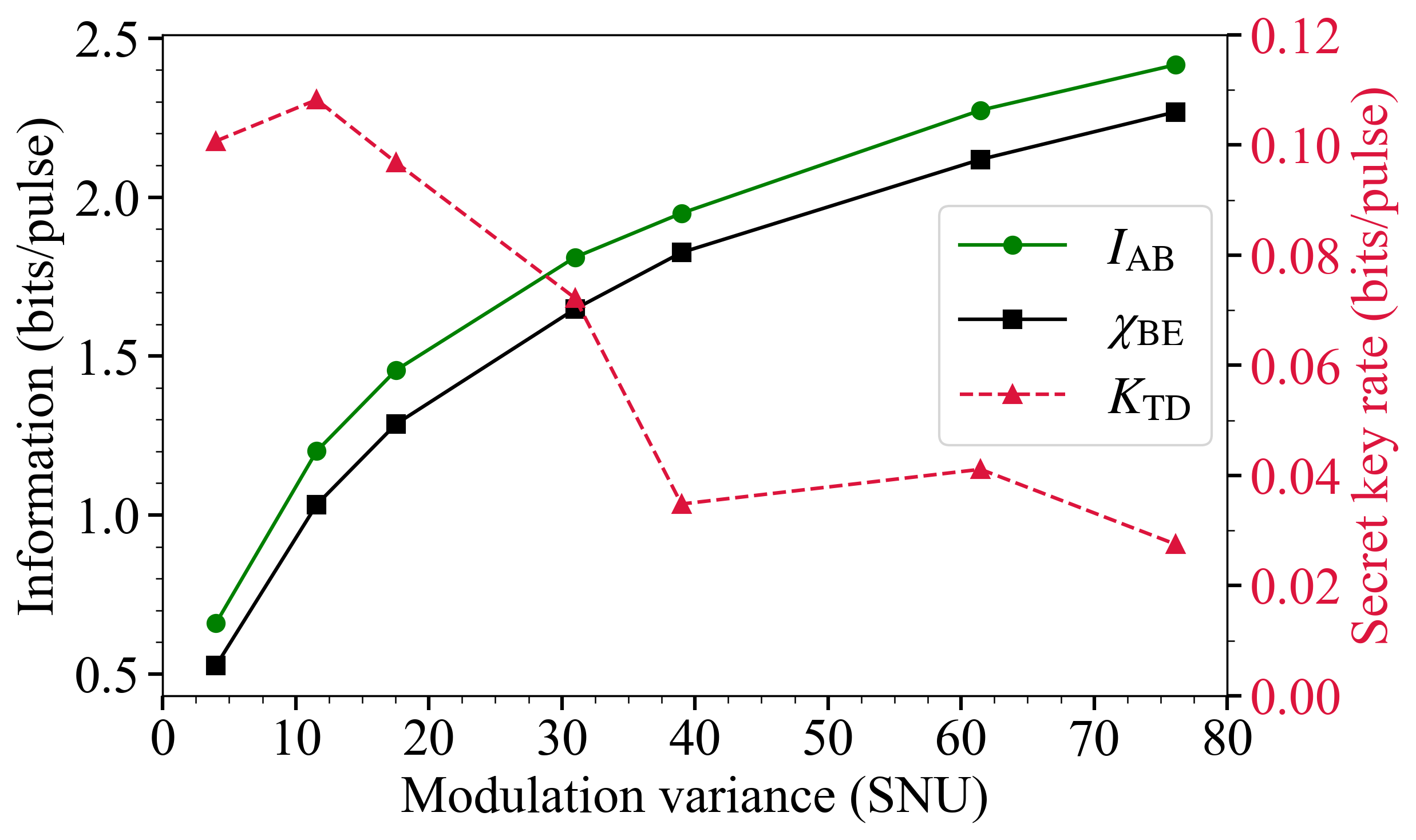}
    \caption{Dependence of the TD secret key rate $K_\mathrm{TD}$, mutual information $I_{\mathrm{AB}}$, and Holevo bound $\chi_{\mathrm{BE}}$ on Alice’s modulation variance $V_A$. The key rate increases with $V_A$ at low modulation, reaches an optimum, and decreases at higher values due to the associated reduction in effective channel transmittance, limiting secure key generation.}
    \label{fig:keyrate}
\end{figure}

Figure~\ref{fig:keyrate} presents the dependence of $K_{\mathrm{TD}}$, $I_{\mathrm{AB}}$, and $\chi_{\mathrm{BE}}$ on $V_A$, using the parameters in Table~\ref{tab:udcvqkd_params}. Both $I_{\mathrm{AB}}$ and $\chi_{\mathrm{BE}}$ increase with $V_A$, in qualitative agreement with the numerical results shown in Figure~\ref{fig:K-VA}. Theoretically, $K_{\mathrm{TD}}$ exhibits a maximum at an optimal modulation variance $V_A^{\mathrm{opt}}=23.63$ and then decreases as the growth of $\chi_{\mathrm{BE}}$ surpasses $\beta I_{\mathrm{AB}}$ for $\beta < 1$. Experimentally, within the accessible low-$V_A$ range, $K_{\mathrm{TD}}$ increases from $V_A=4.01$ to $V_A=11.57$ SNU, where it reaches a maximum of approximately $0.1082$ bits/pulse (corresponding to $\sim 270$ kbps at a repetition rate of $2.5$ MHz) at $V_A^{\mathrm{opt}}=11.57$ SNU, and subsequently decreases with further increase in $V_A$. However, the observed degradation of $K_{\mathrm{TD}}$ is more pronounced than numerical prediction, with $K_{\mathrm{TD}}$ decreasing to $\sim 0.0275$ bits/pulse. For sufficiently large modulation variances, $V_A = 91.10, 96.88$, no secure key rate is obtained experimentally (not shown). This degradation is associated with a non-uniform $T_x$, which is higher at lower $V_A$ values ($V_A = 4.01$ to $17.54$ SNU) and comparatively decreases at higher $V_A$ (from $30.98$ SNU onward). In terms of the security analysis, the reduction in $T_x$ effectively shifts the $C_p$-independent secure region toward lower $V_p^B$ threshold (in Figure~\ref{fig:security}(b)), tightening the admissible $V_p^B$ range. When $(V_p^B)_{\max} <1.003$ SNU (which is experimentally measured), the system enters the unsecure regime and the secure key rate vanishes under worst-case assumptions on $C_p$. 

\begin{figure}[h]
    \centering
    \includegraphics[scale = 0.45]{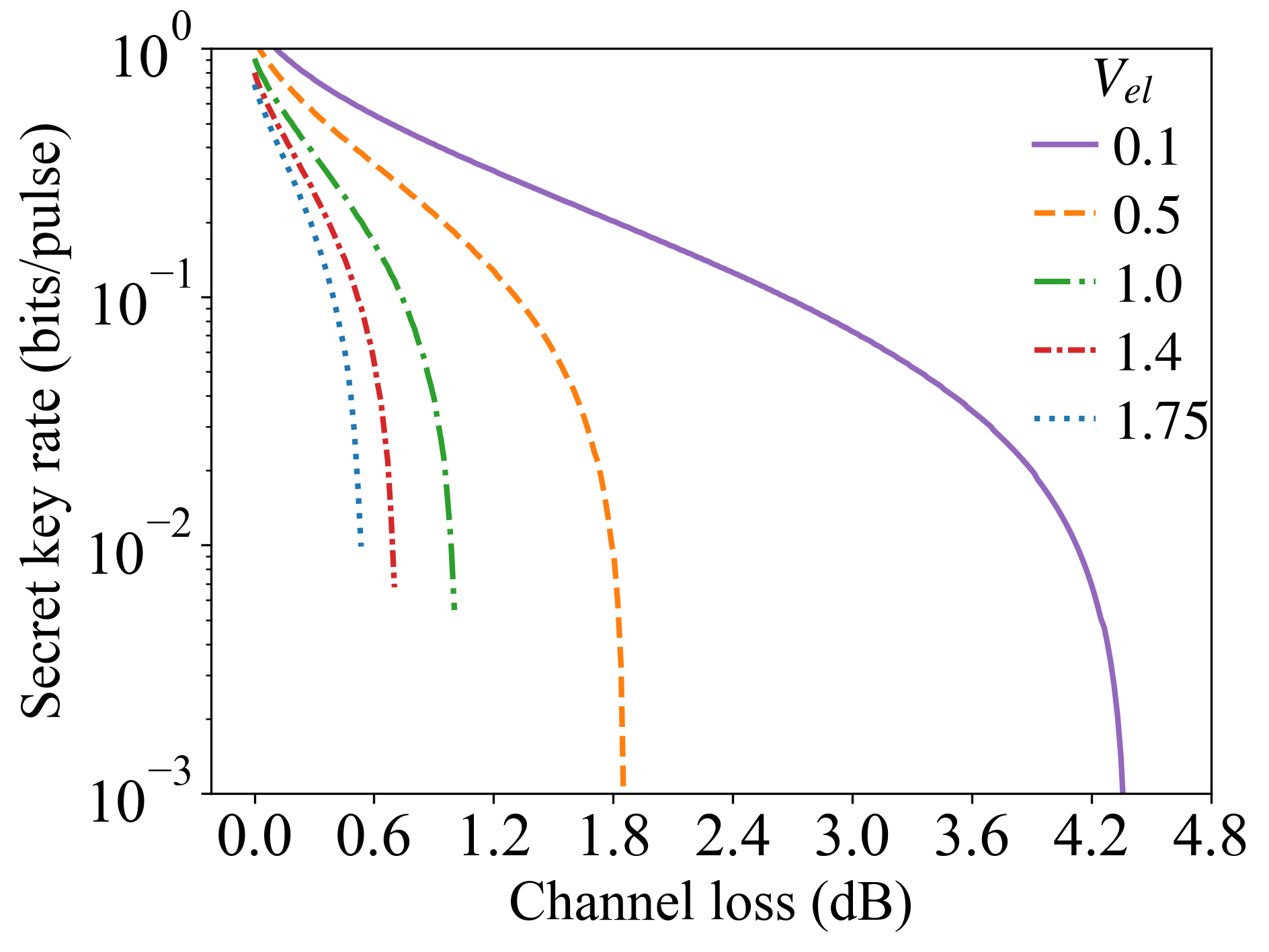}
    \caption{Secret key rate (TD) as a function of channel loss (in dB) for different electronic noise variances $V_{el}$ (in SNU). Each curve exhibits a finite cutoff loss beyond which the key rate becomes zero, defining the security limit. Simulation parameters are $V_A=20$ SNU, $\xi_x = 0.01$ SNU, $\eta = 0.81$, and $V_p^B = 1.003$ SNU.}
    \label{fig:keyloss}
\end{figure}

Since the protocol is sensitive to channel transmittance, which is one of the key factor determining the allowable threshold range of $V_p^B$, we investigate its practicability in terms of channel loss. Figure~\ref{fig:keyloss} presents the numerically estimated secret key rate as a function of channel loss (in dB) for different detector electronic-noise values, $V_{el}$ (SNU), thereby quantifying the tolerable loss and corresponding transmission distance. The performance is analyzed across multiple electronic noise regimes: low-noise ($V_{el}=0.1$), moderate-noise ($V_{el}=0.5, 1.0$), and high-noise ($V_{el}=1.4$ (experimental condition) and $1.75$). In all cases, the secret key rate decreases monotonically with increasing channel loss and approaches zero at a finite cutoff, which defines the maximum tolerable loss for secure key generation. Increasing $V_{el}$ leads to both a reduction in the achievable key rate and a restriction of the secure transmission range. Specifically, the cutoff loss (with the corresponding channel transmittance $T_x$ given in parentheses) decreases from approximately $4.381$ dB ($T_x \approx 0.365$) for $V_{el}=0.1$ to $1.856$ dB ($0.652$), $1.003$ dB ($0.793$), $0.702$ dB ($0.851$), and $0.535$ dB ($0.884$) for $V_{el}=0.5,\ 1.0,\ 1.4,$ and $1.75$, respectively. For our experimental case ($V_{el}=1.4$ SNU), the cutoff transmittance is $T_x^\mathrm{cutoff}=0.851$, below which no secure key can be generated. In the experimental higher $V_A$ regime discussed previously, the measured transmittance falls below this threshold ($T_x < T_x^\mathrm{cutoff}$), consistent with the absence of a secure key rate. Assuming a standard fiber attenuation coefficient of 0.2 dB/km, these cutoff losses correspond to maximum secure transmission distances of approximately  $21.9$ km, $9.3$ km, $5.0$ km, $3.5$ km, and $2.7$ km, respectively. Hence, secure key generation remains feasible with high electronic-noise detectors; however, the achievable transmission distance is limited to low-loss (high-transmission) channels, effectively restricting operation to short-range links. The results further show that increasing electronic noise reduces the secure communication range, while lower-noise detection can extend the attainable transmission distance.    

Overall, the results demonstrate the feasibility of free-space UD-CVQKD under realistic high electronic-noise detector conditions within the detector-trust assumption. Secure key generation is experimentally demonstrated over short laboratory distances, while numerical results indicate its extension to short-to-moderate transmission ranges, with performance primarily limited by channel transmittance and detector noise.    

\section{Conclusion} \label{sec:conclusion}

In this work, we experimentally demonstrate a Gaussian-modulated UD-CVQKD system in a laboratory free-space channel under realistic high detector electronic-noise conditions ($V_{el} \approx 1.4$ SNU) and analyze its security within both TD and UTD assumptions, depending on the unmodulated $P$-quadrature. The worst-case secure key rates are evaluated within the unknown $C_p$-independent security threshold region. While the UTD model cannot guarantee a positive secure key rate under the considered detector noise, the TD model enables secure key generation over a finite range of modulation variances, achieving a maximum key rate of $0.1082$ bits/pulse ($270$ kbps) at $V_A^{\mathrm{opt}} \approx 11.57$ SNU. Numerical analysis shows a significant reduction in tolerable channel loss and operational range, with a cutoff transmittance of $T_x^{\mathrm{cutoff}} \approx 0.85$, thereby restricting secure operation to low-loss channels. These results quantify the impact of detector electronic noise on practical system performance and establish its role in limiting the achievable secure operating regime. Overall, this work demonstrates the feasibility of free-space UD-CVQKD under high-noise detector conditions, with applicability to short- to moderate-distance implementations.

\section*{Acknowledgments}
The authors acknowledge the support of the Department of Space, Government of India.

\section*{Data Availability}
The data that support the findings of this study are available from the corresponding author upon reasonable request.

\section*{Disclosures}
The authors declare that they have no conflicts of interest related to this article.

\bibliography{Manuscript}

\end{document}